\documentclass{article}
\usepackage[a4paper,margin=1in]{geometry}
\usepackage{amsmath,amssymb}
\usepackage{graphicx}
\usepackage{bm}
\usepackage{siunitx}
\usepackage{hyperref}
\usepackage{xcolor}
\usepackage{cite}
\usepackage{setspace}

\title{\bf Competition between Charge Density Wave and Superconductivity in a Janus MXene Mo$_2$NF$_2$}

\author{
Jakkapat Seeyangnok$^{1a}$,
Udomsilp Pinsook$^{1b}$,
Graeme J Ackland$^{2c}$\\[1ex]
$^{1}$Department of Physics, Faculty of Science, \\
Chulalongkorn University, Bangkok, Thailand\\
$^{2}$Centre for Science at Extreme Conditions,\\
School of Physics and Astronomy,\\
University of Edinburgh, Edinburgh, United Kingdom\\[1ex]
\texttt{$^{a}$jakkapatjpt@gmail.com}\\
\texttt{$^{b}$udomsilp.p@chula.ac.th}\\
\texttt{$^{c}$gjackland@ed.ac.uk}
}

\date{} 

\begin{document}

\maketitle

\begin{abstract}
Charge-density-wave (CDW) order and superconductivity often compete in low-dimensional materials, yet their interplay in Janus MXenes remains largely unexplored. Here, we present a comprehensive first-principles investigation of the structural, vibrational, and electronic properties of Mo$2$NF$2$. Phonon calculations reveal an unstable soft phonon mode at the $M$ point in the high-symmetry structure, signaling a CDW instability. Analysis of phonon linewidths and the real and imaginary parts of the bare electronic susceptibility demonstrates that the CDW is not driven by simple Fermi-surface nesting but instead originates from strong momentum-dependent electron–phonon coupling. Structural relaxation yields a commensurate CDW phase characterized by bond-length modulations involving the Mo, N, and F sublattices. We further show that charge doping alone is insufficient to stabilize the soft phonon, whereas compressive biaxial strain exceeding $-3\%$ completely suppresses the CDW instability. Electron–phonon coupling calculations indicate that the CDW phase exhibits a reduced coupling constant $\lambda = 0.40$ and logarithmic phonon frequency $\omega{\log} = 219$ K, leading to a low superconducting transition temperature of $T_c \sim 1$ K. In contrast, the strain-stabilized high-symmetry phase shows enhanced coupling ($\lambda = 0.53$, $\omega{_{log}} = 272$ K) and a higher $T_c \sim 4$ K. Our results establish Mo$_2$NF$_2$ as a strain-tunable platform where superconductivity emerges upon suppression of a competing CDW phase, highlighting the crucial role of lattice control in Janus MXenes.
\end{abstract}


\section{Introduction}
Charge-density-wave (CDW) order represents a fundamental manifestation of collective behavior in low-dimensional solids, arising from the coupled reorganization of electronic states and the crystal lattice. A CDW state is characterized by a periodic modulation of the electronic charge density that is accompanied by a symmetry-breaking lattice distortion, reflecting the intimate interplay between electrons and phonons in reduced dimensions~\cite{gruner2018density}. The earliest theoretical description of CDWs was provided by Peierls in the context of one-dimensional metallic systems, where perfect nesting of the Fermi surface produces a divergence in the electronic susceptibility and drives a lattice instability at a characteristic wave vector~\cite{peierls1955quantum}. This electronic instability is inherently linked to phonon softening via electron–phonon coupling, a mechanism later formalized through the concept of the Kohn anomaly~\cite{kohn1959image}.

While the Peierls framework provides an elegant description of CDW formation in idealized one-dimensional systems, it has become increasingly clear that this picture is insufficient to explain CDW phenomena in higher dimensions. In quasi-two-dimensional materials, the Fermi surface typically lacks the perfect nesting required to sustain a purely electronic instability. Instead, both experimental observations and first-principles calculations have demonstrated that CDW transitions in realistic materials are governed by a cooperative interaction between electronic states and lattice degrees of freedom, mediated by a strongly momentum-dependent electron–phonon coupling~\cite{johannes2008fermi,zhu2017misconceptions}. In this scenario, selective softening of specific phonon modes at well-defined wave vectors plays a decisive role, and the CDW transition is best understood as a coupled electronic–structural instability rather than a purely electronic Peierls transition~\cite{zhu2017misconceptions}.

Layered transition-metal dichalcogenides (TMDs) provide a paradigmatic class of materials for investigating CDW physics in two dimensions. Prototypical compounds such as NbSe$_2$~\cite{silva2016electronic,weber2011extended,ugeda2016characterization,nakata2021robust}, TaSe$_2$~\cite{johannes2008fermi,nakata2021robust,xi2015strongly}, and TaS$_2$~\cite{dalal2025flat,tsen2015structure,philip2023local} exhibit robust CDW phases despite the absence of pronounced Fermi-surface nesting. In these systems, detailed momentum-resolved studies have established that the CDW instability originates from highly anisotropic electron–phonon coupling, which selectively enhances phonon softening at specific wave vectors and stabilizes symmetry-lowering lattice distortions~\cite{weber2011extended,calandra2011charge}. The weak interlayer coupling in TMDs further enables the persistence of CDW order down to the monolayer limit, where genuinely two-dimensional CDW phases have been experimentally realized~\cite{ugeda2016characterization,xi2015strongly}. In hydrogen-functionalized TMDs, MoSH has been proposed to exhibit stronger lattice instabilities in the 1T phase, which may be associated with charge-density-wave (CDW) formation~\cite{ku2023ab} and in its bilayer configuration~\cite{noor2025superconductivity}. More recently, signatures of CDW ordering have been observed in both the 2H and 1T phases of MoSeH~\cite{sui2025two}. Charge-density-wave (CDW) phases in both MoSH and MoSeH have recently been examined in detail~\cite{seeyangnok2026moxhcdw}. A similar interplay between superconductivity and CDW distortions has been suggested for hydrogen-functionalized Janus systems such as WSH and WSeH, where superconductivity with Tc values on the order of 10-15 K coexists with structural modulations characteristic of CDW ordering~\cite{seeyangnok2024superconductivity,seeyangnok2024superconductivitywseh,qiao2024prediction}. Furthermore, several Janus group-IV transition-metal MXH monolayers have been theoretically investigated and predicted to host phonon-mediated superconductivity with Tc ranging from 9 to 30 K~\cite{li2024machine,ul2024superconductivity,seeyangnok2025competition}. Importantly, some of these hydrogen-functionalized systems also display lattice instabilities indicative of competing CDW phases, similar to those identified in MoSH-, WSH-, and WSeH-based monolayers, highlighting the delicate balance between superconductivity and structural ordering in this class of materials. As a result, TMDs have emerged as an ideal platform for exploring CDW formation beyond one dimension and for examining its competition and coexistence with other collective phenomena, most notably superconductivity. 

In contrast to other halogen-functionalized systems, halogen functionalization in Mo$_2$C has been shown to enhance superconductivity~\cite{seeyangnok2026mo2c}. In this work, we present a comprehensive first-principles study of the charge-density-wave (CDW) instability and its interplay with superconductivity in Mo$_2$NF$_2$. By combining phonon dispersion, phonon linewidth analysis, and momentum-resolved electronic susceptibility, we identify a soft phonon mode at the $M$ point as the microscopic origin of the CDW instability, driven by strong momentum-dependent electron-phonon coupling rather than Fermi-surface nesting. We further resolve the real-space CDW structure and reveal cooperative lattice distortions of the Mo, N, and F sublattices.Moreover, we show that external perturbations affect the instability differently: charge doping fails to stabilize the soft mode, whereas moderate compressive biaxial strain suppresses the CDW, restores the high-symmetry phase, and enhances electron-phonon coupling, significantly increasing the superconducting transition temperature. These results establish strain as an effective parameter for tuning competing collective phases in Janus MXenes and deepen the understanding of lattice-driven CDW physics in two-dimensional materials.
\section{Computational Methods}
First-principles calculations were performed within density functional theory (DFT) as implemented in the \textsc{Quantum ESPRESSO} package~\cite{giannozzi2009quantum}. The exchange--correlation functional was treated using the generalized gradient approximation in the Perdew--Burke--Ernzerhof (GGA-PBE) form~\cite{perdew1996generalized}. Optimized norm-conserving Vanderbilt pseudopotentials~\cite{schlipf2015optimization} were employed for all atomic species. The kinetic-energy cutoffs for the plane-wave basis set and charge density were set to 80~Ry and 320~Ry, respectively.

Structural optimizations were carried out by fully relaxing atomic positions and lattice parameters until the residual forces on each atom were below $10^{-5}$~eV/\AA. Brillouin-zone integrations for self-consistent calculations were performed using Monkhorst--Pack $\mathbf{k}$-point meshes~\cite{monkhorst1976special} of $24\times24\times1$ for the primitive unit cell, together with Methfessel--Paxton smearing~\cite{methfessel1989high} of width 0.02~Ry. For supercell calculations, the $\mathbf{k}$-point meshes were reduced proportionally to the supercell size, corresponding to $12\times12\times1$, $9\times9\times1$, $6\times6\times1$, and $3\times3\times1$ meshes for $2\times2\times1$, $3\times3\times1$, $4\times4\times1$, and $5\times5\times1$ supercells, respectively.

Several magnetic configurations were examined to determine the magnetic ground state of Mo$_2$NF$_2$ MXenes. These include a ferromagnetic (FM) configuration, in which the initial magnetic moments of all Mo atoms were aligned parallel, as well as antiferromagnetic (AFM) configurations. The AFM states considered consist of a G-type antiferromagnetic (GAF) ordering, characterized by antiparallel alignment between neighboring Mo atoms, and an A-type antiferromagnetic (AAF) ordering, where spins are aligned ferromagnetically within the lower Mo plane but antiparallel to those in the upper Mo plane. The calculations indicate that none of the magnetic configurations stabilize a magnetic insulating phase. Instead, Mo$_2$NF$_2$ MXenes favor a metallic ground state regardless of the initial magnetic ordering. This behavior suggests that itinerant electronic states dominate near the Fermi level, suppressing long-range magnetic ordering in the high-symmetry phase.

Phonon dispersions and lattice dynamical properties were calculated within density functional perturbation theory (DFPT)~\cite{baroni2001phonons}. Dynamical matrices were computed on uniform $12\times12\times1$ $\mathbf{q}$-point meshes for the primitive unit cell and on $6\times6\times1$ meshes for the $2\times2\times1$ CDW supercells. Dynamical stability was assessed by inspecting the phonon spectra for imaginary frequencies. Electron--phonon interactions were evaluated within the density functional perturbation theory (DFPT) framework~\cite{baroni2001phonons}. 

The superconducting transition temperature $T_c$ was evaluated within the Allen--Dynes formalism using the isotropic Eliashberg spectral function $\alpha^2F(\omega)$~\cite{allen1975transition,pinsook2024analytic}. It is given by
\begin{equation}
T_c = \frac{\omega_{\ln}}{1.2}
\exp\left[
-\frac{1.04(1+\lambda)}
{\lambda - \mu^{\ast}(1+0.62\lambda)}
\right],
\end{equation}
where $\lambda$ denotes the electron--phonon coupling (EPC) constant and $\mu^{\ast}$ is the Coulomb pseudopotential.

The EPC constant $\lambda$ is obtained from the Eliashberg spectral function as
\begin{equation}
\lambda = 2 \int_0^{\infty} \frac{\alpha^2F(\omega)}{\omega} \, d\omega.
\end{equation}

The logarithmic average phonon frequency $\omega_{\ln}$, which characterizes the effective phonon energy scale, is defined as
\begin{equation}
\omega_{\ln} =
\exp\left[
\frac{2}{\lambda}
\int_0^{\infty}
\frac{\alpha^2F(\omega)}{\omega}
\ln \omega \, d\omega
\right].
\end{equation}

In addition, the second moment of the phonon spectrum $\omega_2$ is expressed as
\begin{equation}
\omega_2 =
\left[
\frac{2}{\lambda}
\int_0^{\infty}
\alpha^2F(\omega)\,\omega \, d\omega
\right]^{1/2}.
\end{equation}

\section{Results and Discussion}
\subsection{High-symmetry structure (HSS)}
\begin{figure}[h!]
    \centering
    \includegraphics[width=12cm]{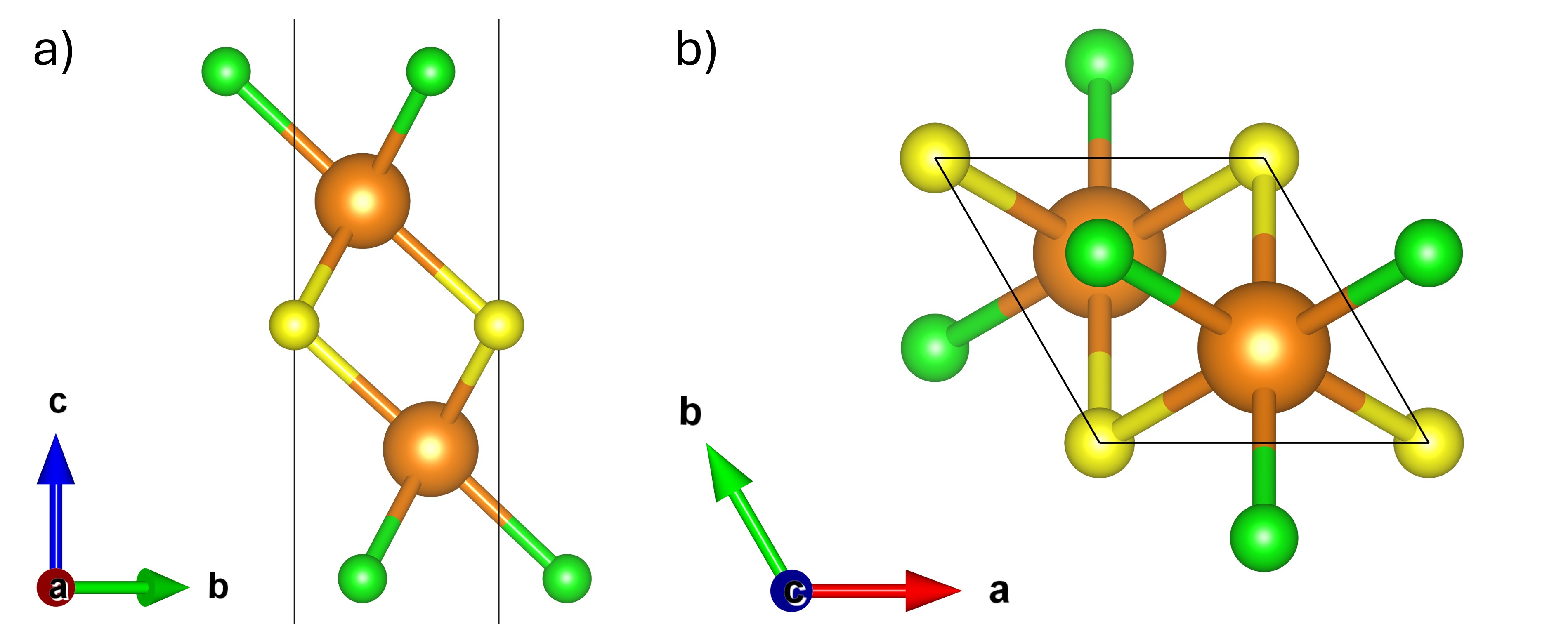}
    \caption{(a) Side and (b) top views of the optimized crystal structure of the Mo$_2$NF monolayer. Orange, yellow, and green spheres represent Mo, N, and F atoms, respectively.}
    \label{fig:structure_monf}
\end{figure}

The high-symmetry phase of Mo$_2$NF$_2$ MXene was investigated within a hexagonal lattice framework. The optimized structure crystallizes in the centrosymmetric space group $P\overline{3}m1$ (No.~164) with an extended Bravais lattice symbol of hP2. To model the two-dimensional geometry, a vacuum spacing of 20.0~\AA was introduced along the out-of-plane ($c$) direction to eliminate spurious interactions between periodic images. In this structure, the nitrogen atom occupies the $1b$ Wyckoff position at $(0,0,0.5)$, forming the central atomic layer. The Mo atoms are located at the $2d$ Wyckoff positions with fractional coordinates $(2/3,1/3,z)$ and $(1/3,2/3,1-z)$, where $z = 0.5726$, resulting in two Mo layers symmetrically displaced about the N plane. The F surface terminations also occupy the $2d$ Wyckoff sites at $(1/3,2/3,z_{\mathrm{F}})$ and $(2/3,1/3,1-z_{\mathrm{F}})$ with $z_{\mathrm{F}} = 0.6484$, leading to symmetric functionalization on both sides of the MXene sheet. The optimized in-plane lattice constant is $a = 2.76$~\AA. Upon fluorine functionalization, the system becomes thermodynamically more stable, with a binding energy of $-9.41$~eV/formula, and exhibits a metallic ground-state configuration.

\begin{figure}[h!]
    \centering
    \includegraphics[width=12cm]{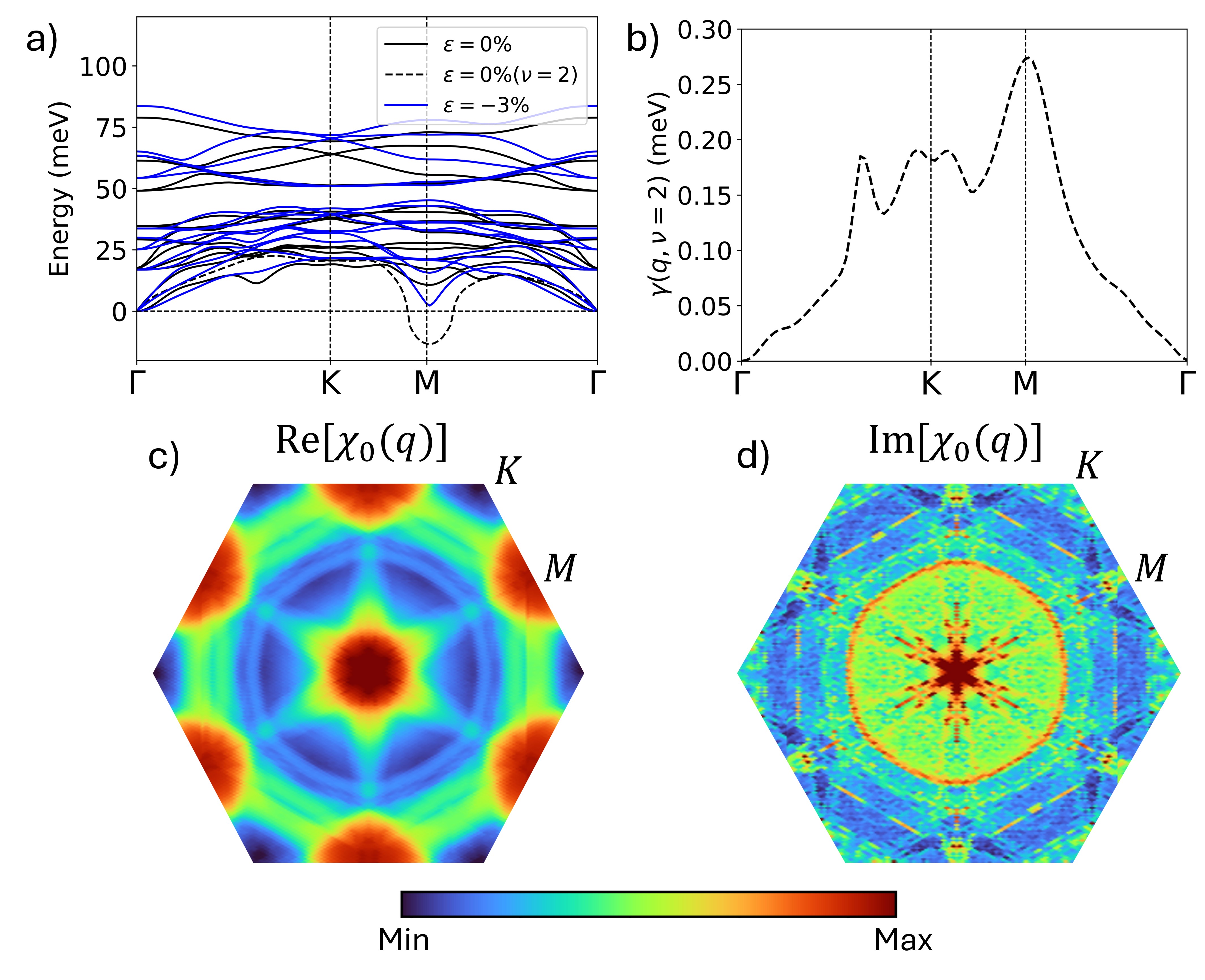}
    \caption{(a) Phonon dispersion along the high-symmetry path $\Gamma$--$K$--$M$--$\Gamma$.
    The black solid lines correspond to the unstrained structure ($\varepsilon = 0\%$),
    while the black dashed line ($\nu = 2$) indicates an unstable soft phonon mode at the
    $M$ point. The blue curves show the phonon dispersion under compressive strain
    ($\varepsilon = -3\%$), where the soft mode is fully stabilized.
    (b) Phonon linewidth $\gamma(\mathbf{q}, \nu = 2)$, exhibiting a pronounced enhancement
    near the $M$ point, indicative of strong electron--phonon coupling.
    (c) Real part of the static electronic susceptibility, $\mathrm{Re}[\chi_{0}(\mathbf{q})]$,
    showing enhanced responses at high-symmetry wave vectors.
    (d) Imaginary part of the electronic susceptibility, $\mathrm{Im}[\chi_{0}(\mathbf{q})]$,
    revealing pronounced features associated with Fermi-surface nesting.}
    \label{fig:phonon_susceptibility}
\end{figure}

To elucidate the microscopic origin of the charge-density-wave (CDW) instability in Fluorine-Functionalized Mo$_2$N, we analyze the phonon dispersions, phonon linewidths, and the real and imaginary parts of the bare electronic susceptibility, as summarized in Fig.~\ref{fig:phonon_susceptibility}. This combined analysis enables us to distinguish between a CDW driven by Fermi-surface nesting and one originating from momentum-dependent electron--phonon coupling. The bare (Lindhard) electronic susceptibility is given by
\begin{equation}
\mathrm{Re}\,[\chi(\mathbf{q},\omega)] =
\sum_{\mathbf{k}}
\frac{f(\varepsilon_{\mathbf{k}}) - f(\varepsilon_{\mathbf{k}+\mathbf{q}})}
{\omega + \varepsilon_{\mathbf{k}} - \varepsilon_{\mathbf{k}+\mathbf{q}}},
\end{equation}
\begin{equation}
\mathrm{Im}\,[\chi(\mathbf{q},\omega)] =
\pi \sum_{\mathbf{k}}
\left[f(\varepsilon_{\mathbf{k}}) - f(\varepsilon_{\mathbf{k}+\mathbf{q}})\right]
\delta(\omega + \varepsilon_{\mathbf{k}} - \varepsilon_{\mathbf{k}+\mathbf{q}}),
\end{equation}
where $\varepsilon_{\mathbf{k}}$ denotes the electronic band energy and $f(\varepsilon)$ is the Fermi--Dirac distribution function.

In the static limit ($\omega \rightarrow 0$), the imaginary part, Im$[\chi_0(\mathbf{q})] = \mathrm{Im}[\chi(\mathbf{q},0)]$, primarily reflects the topology of the Fermi surface and is often used as an indicator of Fermi-surface nesting. In contrast, the real part, Re$[\chi_0(\mathbf{q})] = \mathrm{Re}[\chi(\mathbf{q},0)]$, governs the stability of the electronic system and directly contributes to the renormalization of phonon frequencies. A pronounced enhancement in Re$[\chi_0(\mathbf{q})]$ at a particular wave vector can induce phonon softening and drive a structural instability. Consequently, a CDW of purely electronic (Peierls-type) origin requires coincident peaks in both Re$[\chi_0(\mathbf{q})]$ and Im$[\chi_0(\mathbf{q})]$ at the same ordering vector $\mathbf{q}_{\mathrm{CDW}}$.

As shown in Fig.~\ref{fig:phonon_susceptibility}(a), the phonon spectrum of unstrained Mo$_2$NF$_2$ exhibits a pronounced softening of a low-energy phonon branch at the $M$ point, signaling a lattice instability toward a commensurate modulation with $\mathbf{q}_{\mathrm{CDW}} \approx M$. The corresponding phonon linewidth, shown in Fig.~\ref{fig:phonon_susceptibility}(b), displays a strong enhancement near the $M$ point, indicating substantial electron--phonon coupling for this specific phonon mode. The momentum dependence of the electronic susceptibility provides further insight into the origin of this instability. The imaginary part, Im$[\chi_0(\mathbf{q})]$, shown in Fig.~\ref{fig:phonon_susceptibility}(d), exhibits enhanced intensity over extended regions of the Brillouin zone, reflecting the overall geometry of the Fermi surface. However, its maxima are broadly distributed and do not uniquely coincide with the $M$ point. In contrast, the real part, Re$[\chi_0(\mathbf{q})]$, presented in Fig.~\ref{fig:phonon_susceptibility}(c), shows pronounced peaks near the $M$ point, consistent with the momentum location of the soft phonon mode.

The lack of a one-to-one correspondence between the peak positions of Re$[\chi_0(\mathbf{q})]$ and Im$[\chi_0(\mathbf{q})]$ indicates that simple Fermi-surface nesting is not the dominant driving mechanism for the CDW instability in Mo$_2$NF$_2$. Instead, the coincidence of the phonon softening, the enhancement of Re$[\chi_0(\mathbf{q})]$, and the large phonon linewidth at the $M$ point points to a CDW mechanism governed by strong momentum-dependent electron--phonon coupling, rather than a purely electronic Peierls-type instability.


\begin{figure}[h!]
    \centering
    \includegraphics[width=8cm]{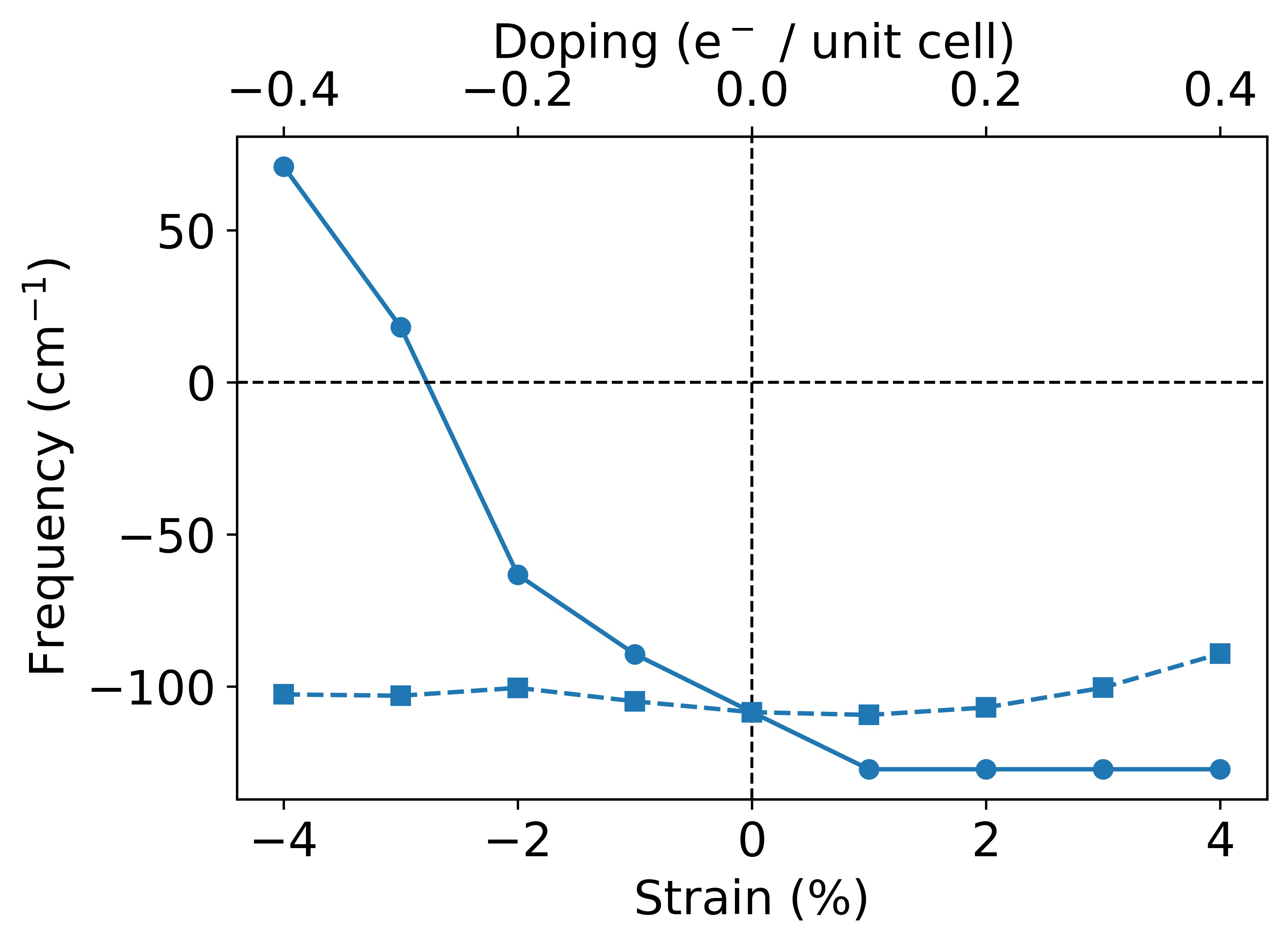}
    \caption{Evolution of the phonon frequency of the unstable mode at the $M$ point as a function
    of biaxial strain and charge doping.
    The bottom horizontal axis shows the applied strain, while the top axis indicates the
    corresponding electron (positive) and hole (negative) doping per unit cell.}
    \label{fig:strain_doping_M}
\end{figure}

External perturbations provide an effective means to control lattice instabilities in low-dimensional materials. To assess possible routes for stabilizing the high-symmetry phase of Mo$_2$NF$_2$, we examine the effects of biaxial strain and charge doping on the unstable soft phonon mode at the $M$ point. In the unperturbed structure, the phonon spectrum exhibits an imaginary-frequency mode at the $M$ point, indicating a dynamical lattice instability. Figure~\ref{fig:strain_doping_M} summarizes the evolution of the phonon frequency of this mode under both biaxial strain and electron or hole doping. Negative phonon frequencies correspond to imaginary modes and reflect an unstable lattice configuration.

Charge doping, introduced by adding or removing electrons from the unit cell, has only a weak influence on the unstable phonon mode. As shown in Fig.~\ref{fig:strain_doping_M}, both electron and hole doping shift the phonon frequency marginally, but the mode remains imaginary over the entire doping range considered. This result indicates that modifying the electronic filling alone is insufficient to suppress the lattice instability. In contrast, biaxial strain has a pronounced effect on the soft phonon. Compressive strain progressively hardens the unstable mode, reducing the magnitude of the imaginary frequency. When the compressive strain exceeds approximately $-3\%$, the phonon frequency at the $M$ point becomes positive, signaling complete stabilization of the soft mode. Tensile strain, on the other hand, further destabilizes the lattice by enhancing the imaginary phonon frequency.

The markedly different responses to strain and doping highlight the dominant role of lattice degrees of freedom in governing the instability. The effectiveness of compressive strain in stabilizing the soft phonon suggests that the instability is primarily controlled by momentum-dependent lattice interactions rather than changes in electronic filling. These results demonstrate that external strain provides a viable and efficient route to suppress the unstable soft phonon and stabilize the high-symmetry phase in Mo$_2$NF$_2$.

\subsection{Charge-Density-Wave (CDW) Structural Distortion}
\begin{figure}[h!]
    \centering
    \includegraphics[width=13cm]{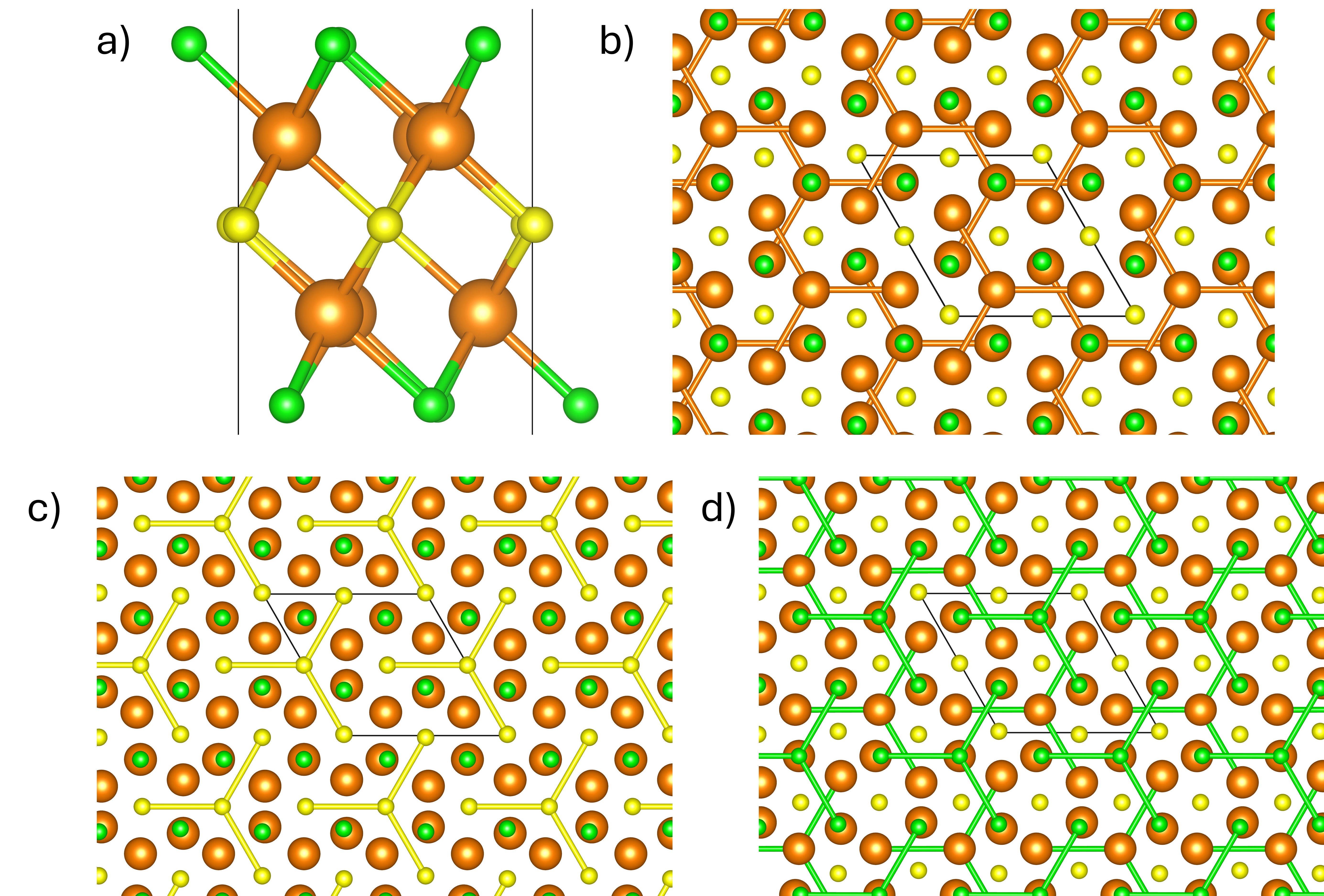}
    \caption{Atomic-scale structural characteristics of the charge-density-wave (CDW) phase in
    Mo$_2$NF$_2$.
    (a) Side view of the CDW structure, showing lattice distortions involving the Mo, N,
    and F sublattices.
    (b) Top view highlighting the shortened Mo--Mo bonds in the CDW phase, with a
    representative bond length of 2.70~\AA.
    (c) Top view illustrating the N--N bond modulation induced by the CDW, with a
    characteristic bond length of 2.72~\AA.
    (d) Top view showing the corresponding F--F bond modulation in the CDW phase, also
    characterized by a bond length of 2.72~\AA.
    Orange, yellow, and green spheres represent Mo, N, and F atoms, respectively.
    }
    \label{fig:cdw_bondlength}
\end{figure}

The relaxed charge-density-wave (CDW) phase of Mo$_2$NF$_2$ exhibits pronounced atomic-scale lattice distortions involving all three atomic sublattices, as illustrated in Fig.~\ref{fig:cdw_bondlength}. In contrast to the high-symmetry structure, the CDW phase is characterized by a periodic modulation of interatomic distances, reflecting the condensation of the unstable soft phonon mode at the $M$ point. As shown in Fig.~\ref{fig:cdw_bondlength}(a), the CDW distortion is not confined to a single atomic layer but instead involves correlated displacements of Mo, N, and F atoms along both in-plane and out-of-plane directions. This collective distortion underscores the strong coupling between the transition-metal framework and the surface terminations, highlighting the cooperative nature of the CDW instability in this Janus MXene. The CDW phase is energetically favored, exhibiting a total energy reduction of
$25.78$~meV/formula compared to the high-symmetry structure and $244.44$~meV/formula compared to the $-3\%$ strain-stabilized high-symmetry phase.

The most prominent structural signature of the CDW appears in the Mo sublattice. Figure~\ref{fig:cdw_bondlength}(b) reveals the formation of shortened Mo--Mo bonds with a representative bond length of 2.70~\AA ~compared to HSS with bond length of 2.76~\AA (Or 2.67~\AA ~when applying -3$\%$ biaxial compressive strain), indicating a partial dimerization of Mo atoms driven by the CDW modulation. This Mo--Mo bond contraction is consistent with the dominant contribution of Mo $d$ states near the Fermi level and reflects the central role of the Mo framework in stabilizing the CDW phase.In addition to the Mo sublattice, the CDW distortion induces measurable bond-length modulations in the anion sublattices. As shown in Fig.~\ref{fig:cdw_bondlength}(c), the N atoms undergo a periodic rearrangement, leading to characteristic N--N bond lengths of approximately 2.72~\AA. A similar modulation is observed for the F surface terminations, as illustrated in Fig.~\ref{fig:cdw_bondlength}(d), where F--F bond lengths of about 2.72~\AA are formed. These distortions indicate that the CDW involves a coherent reconstruction of the entire lattice rather than a purely metal-centered instability.

Overall, the emergence of bond-length alternation across the Mo, N, and F sublattices provides direct real-space evidence of the CDW state in Mo$_2$NF$_2$. The multi-sublattice nature of the distortion, together with the associated phonon softening and enhanced electron--phonon coupling at the $M$ point, confirms that the CDW phase is driven by a lattice instability with strong momentum-dependent coupling, rather than a simple electronic nesting mechanism.

\begin{figure}[h!]
    \centering
    \includegraphics[width=12cm]{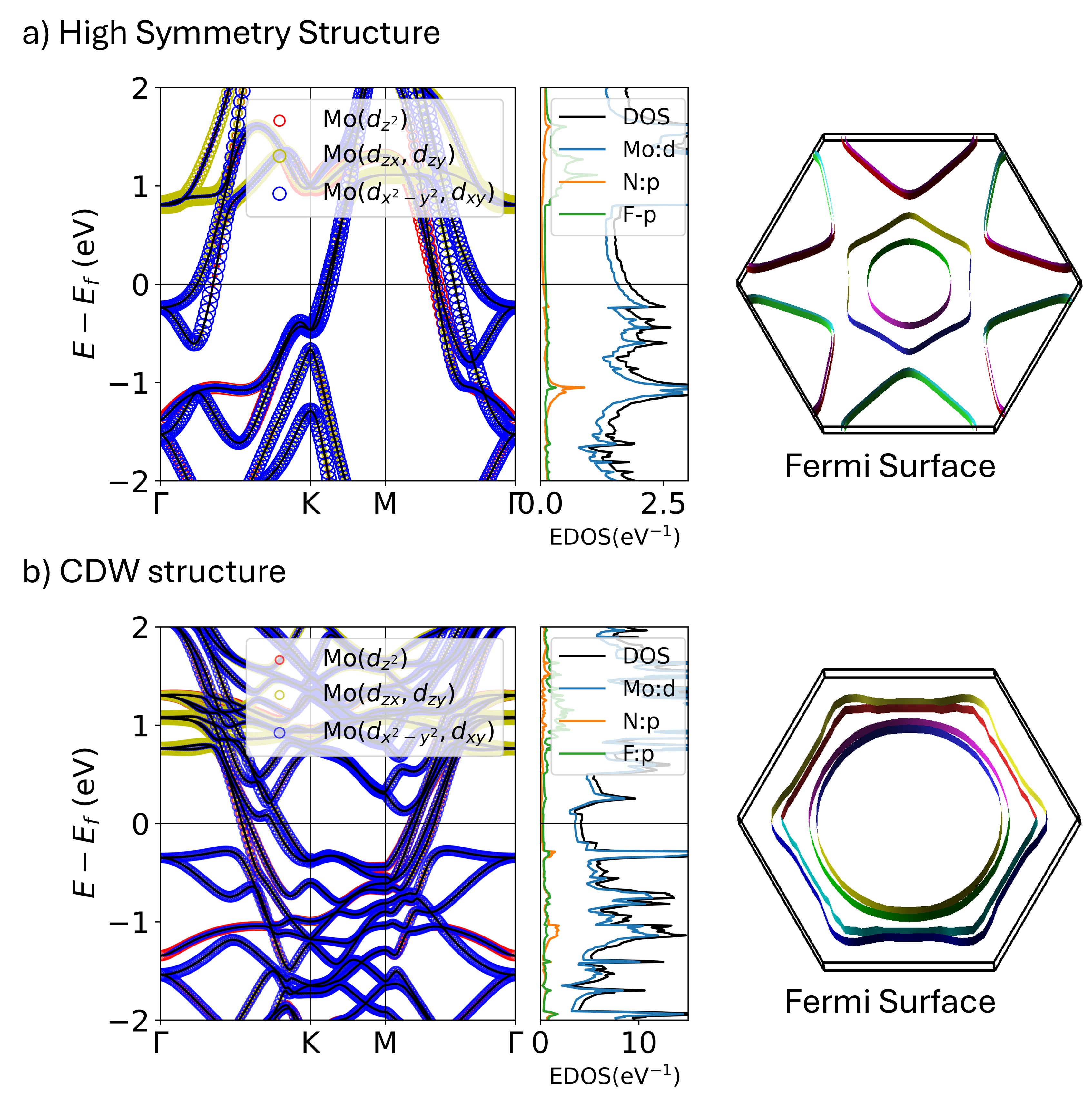}
    \caption{Electronic structure of Mo$_2$NF$_2$ in the high-symmetry and charge-density-wave (CDW) phases.
    (a) High-symmetry structure: orbital-projected electronic band structure along the
    high-symmetry path $\Gamma$--$K$--$M$--$\Gamma$, corresponding electronic density of
    states (EDOS), and Fermi surface.
    (b) CDW structure: orbital-projected band structure, EDOS, and Fermi surface of the
    relaxed CDW phase.
    The bands are colored according to the dominant Mo $d$-orbital contributions
    ($d_{z^2}$, $d_{zx}/d_{zy}$, and $d_{x^2-y^2}/d_{xy}$), while the EDOS is decomposed
    into Mo $d$, N $p$, and F $p$ states.
    The transition from the high-symmetry phase to the CDW phase leads to pronounced band
    reconstruction and a substantial modification of the Fermi surface topology, reflecting
    the lattice distortion and Brillouin-zone folding associated with the CDW ordering.}
    \label{fig:bands_fs_cdw}
\end{figure}

The electronic structures of Mo$2$NF$2$ in the high-symmetry and charge-density-wave (CDW) phases are compared in Fig.~\ref{fig:bands_fs_cdw}. In the high-symmetry structure, the electronic band dispersion exhibits several bands crossing the Fermi level, resulting in a metallic character. Orbital-projected band analysis reveals that the states near the Fermi level are predominantly derived from Mo $d$ orbitals, with the $d{z^2}$ and in-plane $d{x^2-y^2}/d_{xy}$ components providing the largest contributions. The corresponding electronic density of states (EDOS) further confirms the dominance of Mo $d$ states at low energies, while the N $p$ and F $p$ states contribute mainly at deeper binding energies. The Fermi surface of the high-symmetry phase displays hexagonally warped contours centered around the Brillouin-zone center, reflecting the underlying hexagonal lattice symmetry. These Fermi surface features give rise to extended regions with nearly parallel segments, suggesting the presence of partial nesting; however, the nesting is not associated with a single dominant wave vector. This observation is consistent with the absence of a purely electronic instability in the high-symmetry phase and supports the phonon-driven nature of the CDW instability discussed earlier.

Upon relaxation into the CDW phase, the electronic structure undergoes a substantial reconstruction, as shown in Fig.~\ref{fig:bands_fs_cdw}(b). The lattice distortion associated with the CDW leads to Brillouin-zone folding and the appearance of additional bands, resulting in a more complex band dispersion near the Fermi level. Despite this reconstruction, the system remains metallic, with finite density of states at the Fermi energy. The redistribution of spectral weight among the Mo $d$ orbitals indicates a strong coupling between the electronic states and the lattice distortion, particularly within the Mo sublattice. The CDW-induced structural modulation also produces a pronounced reshaping of the Fermi surface. Compared to the high-symmetry phase, the Fermi surface in the CDW phase becomes smoother and more circular, reflecting the reduced symmetry and band folding effects. The absence of a complete Fermi-surface gap further confirms that the CDW transition in Mo$_2$NF$_2$ does not originate from a simple Peierls-type instability. Instead, the observed electronic reconstruction is consistent with a lattice-driven CDW stabilized by strong momentum-dependent electron--phonon coupling, in agreement with the phonon softening and linewidth enhancement at the $M$ point.

\subsection{Competition between CDW and Superconductivity}

\begin{figure}[h!]
    \centering
    \includegraphics[width=14cm]{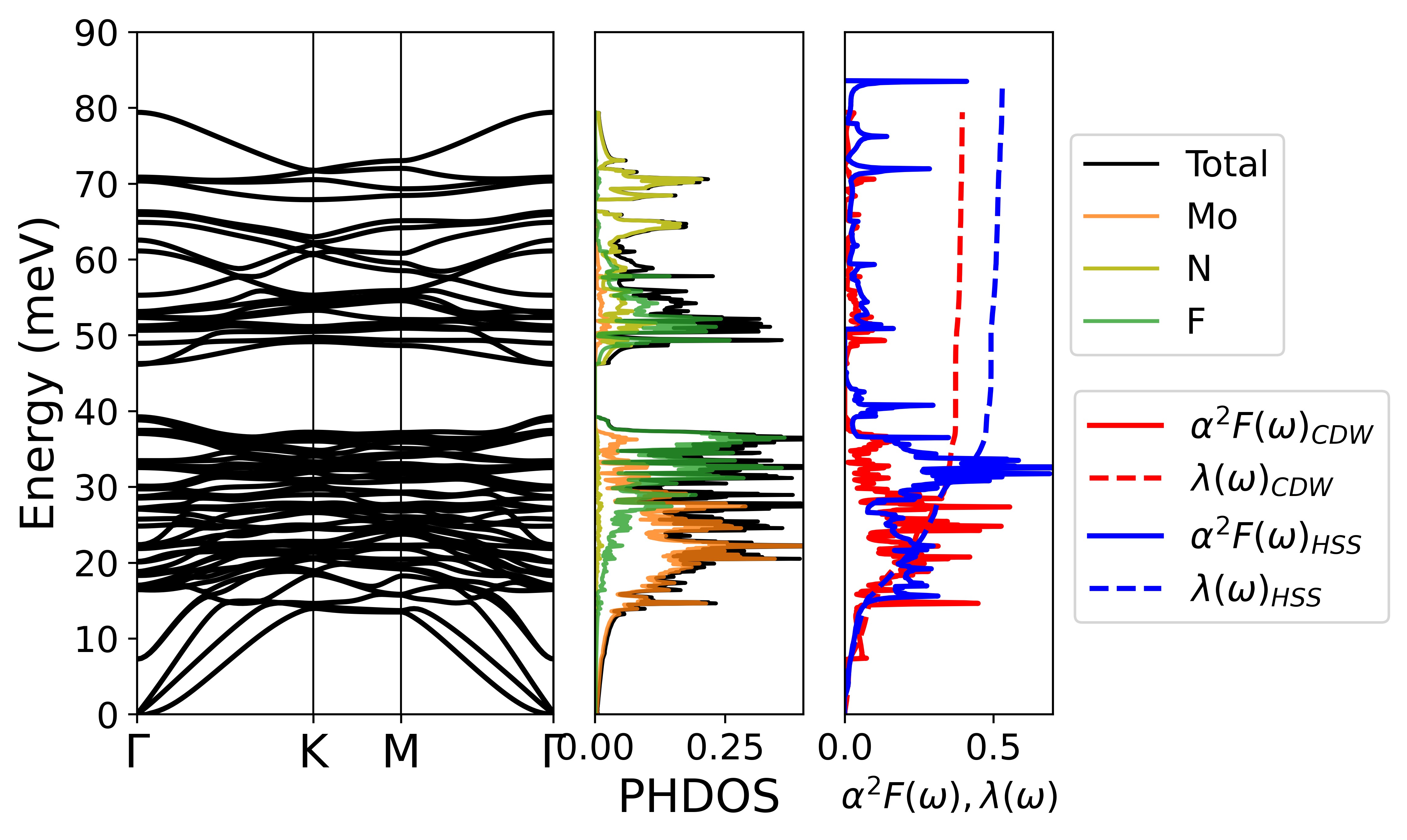}
    \caption{Phonon properties and electron--phonon coupling of Mo$_2$NF$_2$ in the charge-density-wave (CDW) phase and the strain-stabilized high-symmetry structure (HSS). Left panel: phonon dispersion of the relaxed CDW phase along the high-symmetry path $\Gamma$--$K$--$M$--$\Gamma$, showing the absence of imaginary phonon modes and confirming the dynamical stability of the CDW structure. Middle panel: projected phonon density of states (PHDOS), resolved into Mo (orange), N (yellow), and F (green) contributions. Right panel: Eliashberg spectral function $\alpha^2F(\omega)$ (solid lines) and the cumulative electron--phonon coupling constant $\lambda(\omega)$ (dashed lines). Red curves correspond to the CDW phase, yielding a total EPC constant $\lambda = 0.40$ and a logarithmic average phonon frequency $\omega_{\log} = 219$~K. Blue curves represent the high-symmetry structure stabilized under $-3\%$ compressive strain, with an enhanced EPC strength of $\lambda = 0.53$ and $\omega_{\log} = 272$~K.}
    \label{fig:phonon_epc_cdw}
\end{figure}

The electron--phonon coupling (EPC) analysis reveals a clear competition between the charge-density-wave (CDW) phase and superconductivity in Mo$_2$NF$_2$. As summarized in Fig.~\ref{fig:phonon_epc_cdw}, the CDW phase is characterized by a moderate EPC constant of $\lambda = 0.40$ and a logarithmic average phonon frequency $\omega_{\log} = 219$~K, leading to an estimated superconducting transition temperature of $T_c \sim 1$~K. In contrast, the strain-stabilized high-symmetry structure (HSS) under $-3\%$ compressive strain exhibits a significantly enhanced EPC strength of $\lambda = 0.53$ and a higher $\omega_{\log} = 272$~K, resulting in an increased transition temperature of
$T_c \sim 4$~K. The suppression of superconductivity in the CDW phase can be directly traced to the lattice reconstruction and phonon hardening associated with CDW formation. The condensation of the soft phonon mode at the $M$ point stabilizes the CDW structure but simultaneously removes low-energy phonon modes that would otherwise contribute strongly to the EPC. As a result, both the total EPC constant and the characteristic phonon frequency scale are reduced in the CDW phase, diminishing the pairing interaction required for superconductivity. Therefore, Kohn-like phonon softening driven by strong electron-phonon coupling can enhance phonon-mediated superconductivity when it occurs at an optimal level~\cite{jiang2023possible}. However, excessive softening that results in imaginary phonon modes signals the emergence of a charge-density-wave (CDW) instability, which in turn suppresses superconductivity.

Furthermore, the CDW-induced band folding and Fermi-surface reconstruction redistribute electronic spectral weight near the Fermi level, reducing the phase space available for Cooper pairing. Although the CDW phase remains metallic, the partial reconstruction of the electronic structure weakens the coupling between electrons and lattice vibrations, further
suppressing superconductivity. This behavior highlights the antagonistic relationship between CDW order and superconductivity in Mo$_2$NF$_2$.

In contrast, suppressing the CDW instability via compressive strain restores the high-symmetry lattice and stabilizes all phonon modes, thereby enhancing both $\lambda$ and $\omega_{\log}$. The resulting increase in $T_c$ under strain demonstrates that superconductivity in Mo$_2$NF$_2$ is favored in the absence of CDW order and can be effectively tuned through external lattice control. These results establish Mo$_2$NF$_2$ as a representative example of a phonon-mediated superconductor in which superconductivity emerges upon suppression of a competing CDW phase, offering a promising platform for strain-engineered quantum phases in Janus MXenes.

\section{Conclusions}
In conclusion, we have demonstrated that Mo$_2$NF$_2$ hosts a pronounced charge-density-wave instability originating from a soft phonon mode at the $M$ point. Through a combined analysis of phonon dispersions, phonon linewidths, electronic susceptibility, and real-space structural distortions, we establish that the CDW is driven by strong momentum-dependent electron–phonon coupling rather than a purely electronic nesting mechanism. The resulting CDW phase is characterized by cooperative lattice distortions involving the Mo framework as well as the N and F sublattices, confirming the collective nature of the instability in this Janus MXene.

Our results further reveal that external perturbations play markedly different roles in controlling the instability. While charge doping has only a marginal effect and fails to stabilize the soft phonon, moderate compressive biaxial strain efficiently suppresses the CDW by hardening the unstable mode. This strain-induced stabilization of the high-symmetry phase leads to a substantial enhancement of the electron–phonon coupling strength and the characteristic phonon energy scale. Importantly, the CDW phase is shown to suppress superconductivity by reducing both the electron–phonon coupling constant and the logarithmic average phonon frequency, resulting in a low superconducting transition temperature of approximately 1 K. Upon suppression of the CDW under compressive strain, superconductivity is significantly enhanced, with an estimated transition temperature of about 4 K. These findings highlight a clear competition between CDW order and superconductivity in Mo$_2$NF$_2$ and demonstrate that lattice control provides an effective pathway to tune emergent quantum phases.

Overall, this work establishes Janus MXenes as a versatile platform for exploring the interplay between lattice instabilities and superconductivity and suggests that strain engineering may offer a practical route to designing and optimizing superconducting states in two-dimensional materials.

    \section*{Data Availability}
    The data that support the findings of this study are available from the corresponding
    authors upon reasonable request.
    
    \section*{Code Availability}
    The first-principles DFT calculations were performed using the open-source Quantum ESPRESSO package, available at \url{https://www.quantum-espresso.org}, along with pseudopotentials from the Quantum ESPRESSO pseudopotential library at \url{https://pseudopotentials.quantum-espresso.org/}.
\section*{Acknowledgments}
	This research project is supported by the Second Century Fund (C2F), Chulalongkorn University. We acknowledge the supporting computing infrastructure provided by NSTDA, CU, CUAASC, NSRF via PMUB [B05F650021, B37G660013] (Thailand). (\url{URL:www.e-science.in.th}). This also work used the ARCHER2 UK National Supercomputing Service (\url{https://www.archer2.ac.uk}) as part of the UKCP collaboration.

    \section*{Author Contributions}
    Jakkapat Seeyangnok performed all calculations, analyzed the results, wrote the initial draft of the manuscript, and coordinated the project. Graeme J. Ackland and Udomsilp Pinsook contributed to the analysis of the results and to the writing the manuscript.
    
\bibliographystyle{unsrt}
\bibliography{references}

@article{noor2025superconductivity,
  title={Superconductivity and phase stability in various combinations of Janus MoSH bilayers},
  author={Noor ul Taqi, M Munib ul Hassan and Pinsook, Udomsilp and Seeyangnok, Jakkapat},
  journal={Physica Scripta},
  volume={100},
  number={5},
  pages={055939},
  year={2025},
  publisher={IOP Publishing}
}

@misc{seeyangnok2026mo2c,
      title={Enhanced and Tunable Superconductivity Enabled by Mechanically Stable Halogen-Functionalized Mo2C MXenes}, 
      author={Jakkapat Seeyangnok and Udomsilp Pinsook},
      year={2026},
      eprint={2602.11552},
      archivePrefix={arXiv},
      primaryClass={cond-mat.supr-con},
      url={https://arxiv.org/abs/2602.11552}, 
}

@article{jiang2023possible,
  title={Possible enhancement of the superconducting Tc due to sharp Kohn-like soft phonon anomalies},
  author={Jiang, Cunyuan and Beneduce, Enrico and Baggioli, Matteo and Setty, Chandan and Zaccone, Alessio},
  journal={Journal of Physics: Condensed Matter},
  volume={35},
  number={16},
  pages={164003},
  year={2023},
  publisher={IOP Publishing}
}

@article{seeyangnok2026moxhcdw,
  title={Charge Density Wave Order and Superconductivity in Janus MoXH Monolayers},
  author={Seeyangnok, Jakkapat and Pinsook, Udomsilp and Ackland, Graeme J},
  journal={arXiv preprint arXiv:2601.02959},
  year={2026}
}

@article{seeyangnok2024superconductivity,
  title={Superconductivity and electron self-energy in tungsten-sulfur-hydride monolayer},
  author={Seeyangnok, Jakkapat and Ul Hassan, M Munib and Pinsook, Udomsilp and Ackland, Graeme},
  journal={2D Materials},
  volume={11},
  number={2},
  pages={025020},
  year={2024},
  publisher={IOP Publishing}
}

@article{methfessel1989high,
  title={High-precision sampling for Brillouin-zone integration in metals},
  author={Methfessel, MPAT and Paxton, AT},
  journal={physical review B},
  volume={40},
  number={6},
  pages={3616},
  year={1989},
  publisher={APS}
}

@article{seeyangnok2024superconductivitywseh,
  title={Superconductivity and strain-enhanced phase stability of Janus tungsten chalcogenide hydride monolayers},
  author={Seeyangnok, Jakkapat and Pinsook, Udomsilp and Ackland, Graeme J},
  journal={Physical Review B},
  volume={110},
  number={19},
  pages={195408},
  year={2024},
  publisher={APS}
}

@article{pinsook2024analytic,
  title={Analytic solutions of Eliashberg gap equations at superconducting critical temperature},
  author={Pinsook, Udomsilp and Natkunlaphat, Nattawut and Rientong, Komkrit and Tasee, Pakin and Seeyangnok, Jakkapat},
  journal={Physica Scripta},
  volume={99},
  number={6},
  pages={065211},
  year={2024},
  publisher={IOP Publishing}
}

@article{ku2023ab,
  title={Ab initio investigation of charge density wave and superconductivity in two-dimensional Janus 2 H/1 T-MoSH monolayers},
  author={Ku, Ruiqi and Yan, Luo and Si, Jian-Guo and Zhu, Songyuan and Wang, Bao-Tian and Wei, Yadong and Pang, Kaijuan and Li, Weiqi and Zhou, Liujiang},
  journal={Physical Review B},
  volume={107},
  number={6},
  pages={064508},
  year={2023},
  publisher={APS}
}

@article{giannozzi2009quantum,
  title={QUANTUM ESPRESSO: a modular and open-source software project for quantum simulations of materials},
  author={Giannozzi, Paolo and Baroni, Stefano and Bonini, Nicola and Calandra, Matteo and Car, Roberto and Cavazzoni, Carlo and Ceresoli, Davide and Chiarotti, Guido L and Cococcioni, Matteo and Dabo, Ismaila and others},
  journal={Journal of physics: Condensed matter},
  volume={21},
  number={39},
  pages={395502},
  year={2009},
  publisher={IOP Publishing}
}

@article{schlipf2015optimization,
  title={Optimization algorithm for the generation of ONCV pseudopotentials},
  author={Schlipf, Martin and Gygi, Fran{\c{c}}ois},
  journal={Computer Physics Communications},
  volume={196},
  pages={36--44},
  year={2015},
  publisher={Elsevier}
}

@article{monkhorst1976special,
  title={Special points for Brillouin-zone integrations},
  author={Monkhorst, Hendrik J and Pack, James D},
  journal={Physical review B},
  volume={13},
  number={12},
  pages={5188},
  year={1976},
  publisher={APS}
}

@article{perdew1996generalized,
  title={Generalized gradient approximation made simple},
  author={Perdew, John P and Burke, Kieron and Ernzerhof, Matthias},
  journal={Physical review letters},
  volume={77},
  number={18},
  pages={3865},
  year={1996},
  publisher={APS}
}

@article{allen1975transition,
  title={Transition temperature of strong-coupled superconductors reanalyzed},
  author={Allen, Ph B and Dynes, RC},
  journal={Physical Review B},
  volume={12},
  number={3},
  pages={905},
  year={1975},
  publisher={APS}
}

@article{ul2024superconductivity,
  title={Superconductivity in monolayer Janus Titanium-sulfurhydride (TiSH) at ambient pressure},
  author={Noor ul Taqi, M Munib ul Hassan and Pinsook, Udomsilp},
  journal={Journal of Physics: Condensed Matter},
  volume={36},
  number={32},
  pages={325702},
  year={2024},
  publisher={IOP}
}

@article{li2024machine,
  title={Machine learning accelerated discovery of superconducting two-dimensional Janus transition metal sulfhydrates},
  author={Li, Jingyu and Wei, Liuming and Shi, Xianbiao and Shi, Lanting and Si, Jianguo and Liu, Peng-Fei and Wang, Bao-Tian},
  journal={Physical Review B},
  volume={109},
  number={17},
  pages={174516},
  year={2024},
  publisher={APS}
}

@article{seeyangnok2025competition,
  title={Competition between superconductivity and ferromagnetism in 2D Janus MXH (M= Ti, Zr, Hf, X= S, Se, Te) monolayer},
  author={Seeyangnok, Jakkapat and Pinsook, Udomsilp and Ackland, Graeme J},
  journal={Journal of Alloys and Compounds},
  volume={1033},  
  pages={180900},
  year={2025},
  publisher={Elsevier}
}

@book{peierls1955quantum,
  title={Quantum theory of solids},
  author={Peierls, Rudolf Ernst},
  year={1955},
  publisher={Oxford University Press}
}

@article{kohn1959image,
  title={Image of the Fermi Surface in the Vibration Spectrum of a Metal},
  author={Kohn, W},
  journal={Physical Review Letters},
  volume={2},
  number={9},
  pages={393},
  year={1959},
  publisher={APS}
}

@book{gruner2018density,
  title={Density waves in solids},
  author={Gruner, George},
  year={2018},
  publisher={CRC press}
}

@article{johannes2008fermi,
  title={Fermi surface nesting and the origin of charge density waves in metals},
  author={Johannes, MD and Mazin, II},
  journal={Physical Review B—Condensed Matter and Materials Physics},
  volume={77},
  number={16},
  pages={165135},
  year={2008},
  publisher={APS}
}

@article{calandra2011charge,
  title={Charge-density wave and superconducting dome in TiSe 2 from electron-phonon interaction},
  author={Calandra, Matteo and Mauri, Francesco},
  journal={Physical review letters},
  volume={106},
  number={19},
  pages={196406},
  year={2011},
  publisher={APS}
}

@article{nakata2021robust,
  title={Robust charge-density wave strengthened by electron correlations in monolayer 1T-TaSe2 and 1T-NbSe2},
  author={Nakata, Yuki and Sugawara, Katsuaki and Chainani, Ashish and Oka, Hirofumi and Bao, Changhua and Zhou, Shaohua and Chuang, Pei-Yu and Cheng, Cheng-Maw and Kawakami, Tappei and Saruta, Yasuaki and others},
  journal={Nature communications},
  volume={12},
  number={1},
  pages={5873},
  year={2021},
  publisher={Nature Publishing Group UK London}
}

@article{silva2016electronic,
  title={Electronic structure of 2H-NbSe2 single-layers in the CDW state},
  author={Silva-Guill{\'e}n, Jos{\'e} {\'A}ngel and Ordej{\'o}n, Pablo and Guinea, Francisco and Canadell, Enric},
  journal={2D Materials},
  volume={3},
  number={3},
  pages={035028},
  year={2016},
  publisher={IOP Publishing}
}

@article{dalal2025flat,
  title={Flat band physics in the charge-density wave state of 1 T-TaS2 and 1 T-TaSe2},
  author={Dalal, Amir and Ruhman, Jonathan and Venderbos, J{\"o}rn WF},
  journal={npj Quantum Materials},
  volume={10},
  number={1},
  pages={31},
  year={2025},
  publisher={Nature Publishing Group UK London}
}

@article{baroni2001phonons,
  title={Phonons and related crystal properties from density-functional perturbation theory},
  author={Baroni, Stefano and De Gironcoli, Stefano and Dal Corso, Andrea and Giannozzi, Paolo},
  journal={Reviews of modern Physics},
  volume={73},
  number={2},
  pages={515},
  year={2001},
  publisher={APS}
}

@article{tsen2015structure,
  title={Structure and control of charge density waves in two-dimensional 1T-TaS2},
  author={Tsen, Adam W and Hovden, Robert and Wang, Dennis and Kim, Young Duck and Okamoto, Junichi and Spoth, Katherine A and Liu, Yu and Lu, Wenjian and Sun, Yuping and Hone, James C and others},
  journal={Proceedings of the National Academy of Sciences},
  volume={112},
  number={49},
  pages={15054--15059},
  year={2015},
  publisher={National Academy of Sciences}
}

@article{philip2023local,
  title={Local structure anomaly with the charge ordering transition of 1 T-TaS 2},
  author={Philip, Sharon S and Neuefeind, Joerg C and Stone, Matthew B and Louca, Despina},
  journal={Physical Review B},
  volume={107},
  number={18},
  pages={184109},
  year={2023},
  publisher={APS}
}

@article{weber2011extended,
  title={Extended phonon collapse and the origin of the charge-density wave in 2 H-NbSe 2},
  author={Weber, F and Rosenkranz, S and Castellan, J-P and Osborn, R and Hott, R and Heid, R and Bohnen, K-P and Egami, T and Said, AH and Reznik, D},
  journal={Physical review letters},
  volume={107},
  number={10},
  pages={107403},
  year={2011},
  publisher={APS}
}

@article{xi2015strongly,
  title={Strongly enhanced charge-density-wave order in monolayer NbSe2},
  author={Xi, Xiaoxiang and Zhao, Liang and Wang, Zefang and Berger, Helmuth and Forr{\'o}, L{\'a}szl{\'o} and Shan, Jie and Mak, Kin Fai},
  journal={Nature nanotechnology},
  volume={10},
  number={9},
  pages={765--769},
  year={2015},
  publisher={Nature Publishing Group UK London}
}

@article{ugeda2016characterization,
  title={Characterization of collective ground states in single-layer NbSe 2},
  author={Ugeda, Miguel M and Bradley, Aaron J and Zhang, Yi and Onishi, Seita and Chen, Yi and Ruan, Wei and Ojeda-Aristizabal, Claudia and Ryu, Hyejin and Edmonds, Mark T and Tsai, Hsin-Zon and others},
  journal={Nature Physics},
  volume={12},
  number={1},
  pages={92--97},
  year={2016},
  publisher={Nature Publishing Group UK London}
}

@article{zhu2017misconceptions,
  title={Misconceptions associated with the origin of charge density waves},
  author={Zhu, Xuetao and Guo, Jiandong and Zhang, Jiandi and Plummer, EW},
  journal={Advances in Physics: X},
  volume={2},
  number={3},
  pages={622--640},
  year={2017},
  publisher={Taylor \& Francis}
}

@article{sui2025two,
  title={Two-dimensional Janus MoSeH with tunable charge density wave, superconductivity and topological properties},
  author={Sui, Chang-Hao and Qiao, Shu-Xiang and Ding, Hao and Jiang, Kai-Yue and Shang, Shu-Ying and Lu, Hong-Yan},
  journal={Materials Today Physics},
  volume={53},
  pages={101698},
  year={2025},
  publisher={Elsevier}
}

@article{qiao2024prediction,
  title={Prediction of charge density wave, superconductivity and topology properties in two-dimensional Janus 2H/1T-WXH (X= S, Se)},
  author={Qiao, Shu-Xiang and Jiang, Kai-Yue and Sui, Chang-Hao and Xiao, Peng-Cheng and Jiao, Na and Lu, Hong-Yan and Zhang, Ping},
  journal={Materials Today Physics},
  volume={46},
  pages={101485},
  year={2024},
  publisher={Elsevier}
}

\end{document}